\documentclass[aps,pre,twocolumn,groupedaddress,superscriptaddress, showpacs]{revtex4-1}

\usepackage{graphicx}
\usepackage{color}
\usepackage{amsmath}
\usepackage{amsfonts}
\usepackage{amssymb}
\usepackage{dcolumn}
\usepackage[french, english]{babel}
\usepackage{array}

\begin{document}

\title{Shortest path and Schramm-Loewner Evolution}

\author{N. Pos\'e}
  \email{Correspondence and requests for materials should be addressed to N. Pos\'e (posen@ethz.ch)} 
  \affiliation{Computational Physics for Engineering Materials, IfB, ETH Zurich, Wolfgang-Pauli-Strasse 27, CH-8093 Zurich, Switzerland}

\author{K. J. Schrenk}
	\affiliation{Computational Physics for Engineering Materials, IfB, ETH Zurich, Wolfgang-Pauli-Strasse 27, CH-8093 Zurich, Switzerland}
  
\author{N. A. M. Ara\'ujo}
  \affiliation{Computational Physics for Engineering Materials, IfB, ETH Zurich, Wolfgang-Pauli-Strasse 27, CH-8093 Zurich, Switzerland}
  
\author{H. J. Herrmann}
  \affiliation{Computational Physics for Engineering Materials, IfB, ETH Zurich, Wolfgang-Pauli-Strasse 27, CH-8093 Zurich, Switzerland}
  \affiliation{Departamento de F\'isica, Universidade Federal do
Cear\'a, 60451-970 Fortaleza, Cear\'a, Brazil}

\begin{abstract}
We numerically show that the statistical properties of the shortest path on critical percolation clusters are consistent with the ones predicted for Schramm-Loewner evolution (SLE) curves for $\kappa=1.04\pm0.02$. The shortest path results from a global optimization process. To identify it, one needs to explore an entire area. Establishing a relation with SLE permits to generate curves statistically equivalent to the shortest path from a Brownian motion. We numerically analyze the winding angle, the left passage probability, and the driving function of the shortest path and compare them to the distributions predicted for SLE curves with the same fractal dimension. The consistency with SLE opens the possibility of using a solid theoretical framework to describe the shortest path and it raises relevant questions regarding conformal invariance and domain Markov properties, which we also discuss.
\end{abstract}


\maketitle

\pagebreak

Percolation was first introduced by Flory to describe the gelation of polymers \cite{Flory41} and later studied in the context of physics by Broadbent and Hammersley \cite{Broadbent57}. This model is considered the paradigm of connectivity and has been extensively applied in several different contexts, such as, conductor-insulator or superconductor-conductor transitions, flow through porous media, sol-gel transitions, random resistor network, epidemic spreading, and resilience of network-like structures \cite{Wilkinson83,Lenormand89,Stauffer94,Sahimi94,Grassberger83,Cardy85,Cohen00,Schneider11}. In the lattice version, lattice elements (either sites or bonds) are occupied with probability $p$, and a continuous phase transition is observed at a critical probability $p_c$, where for $p<p_c$, as the correlation function decays exponentially, all clusters are of exponentially small size, and for $p> p_c$ there is a spanning cluster. At $p_c$, the spanning cluster is fractal \cite{Essam80}. In this article we focus on the shortest path, defined as the minimum number of lattice elements which belong to the spanning cluster and connect two opposite borders of the lattice \cite{Pike81,Herrmann84}. The shortest path is related with the geometry of the spanning cluster \cite{Coniglio81,Pike81,Herrmann84b,Grassberger99b,Pose12}. Thus, studies of the shortest path resonate in several different fields. For example, the shortest path is used in models of hopping conductivity to compute the decay exponent for superlocalization in fractal objects \cite{Harris87,Aharony90}. It is also considered in the study of flow through porous media to estimate the breakthrough time in oil recovery \cite{Soares04} and to compute the hydraulic path of flows through rock fractures \cite{Wettstein12}. The shortest path has even been analyzed in cold atoms experiments to study the breakdown of superfluidity \cite{Krinner13}. However, despite its relevance, the fractal dimension of the shortest path is among the few critical exponents in two-dimensional percolation that are not known exactly \cite{Grassberger85,Zhou12}.

\begin{figure}[t]
\vspace*{-5mm}
\begin{center}
\includegraphics[width=\columnwidth]{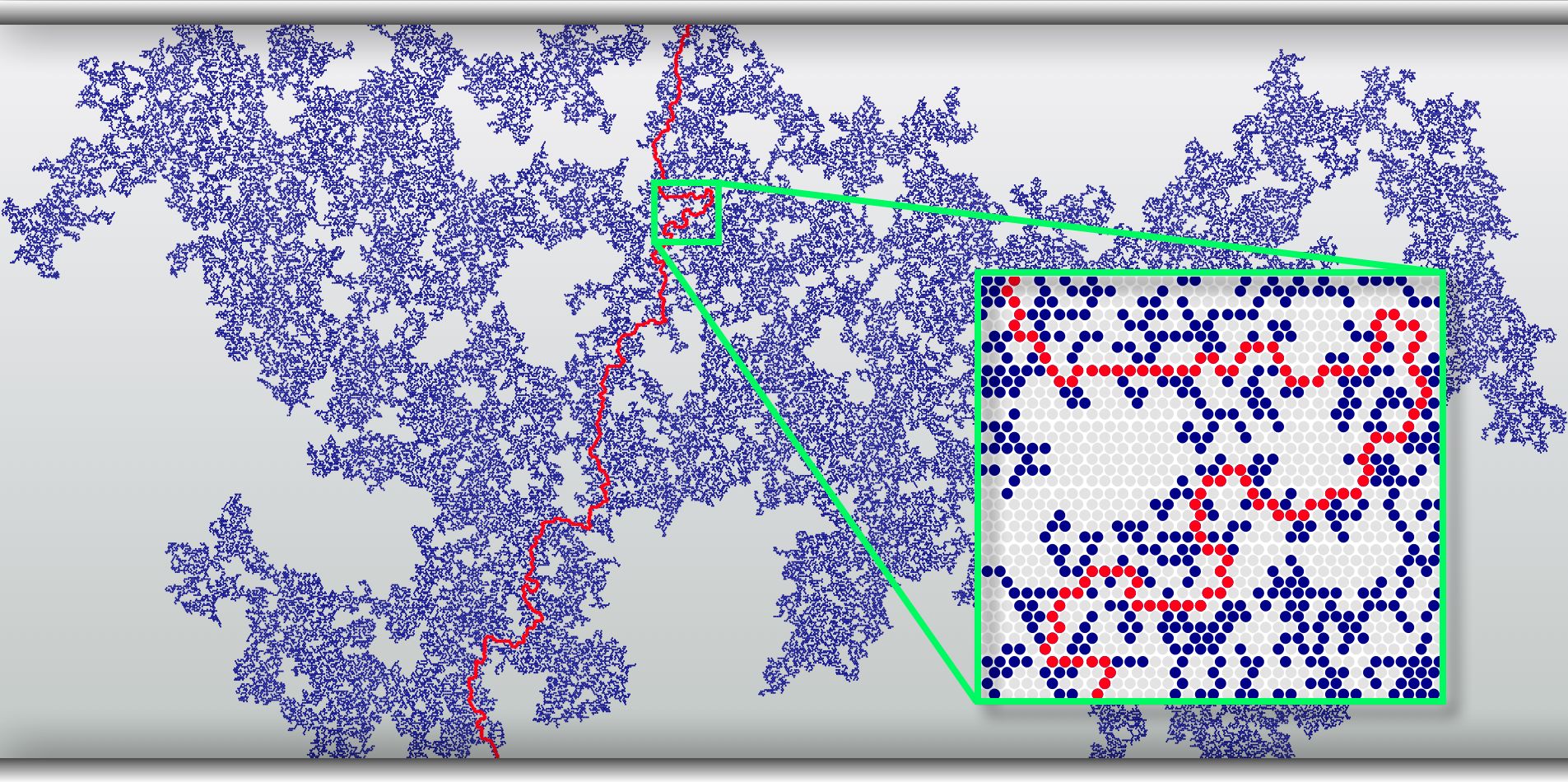}
\end{center}
\vspace*{-3mm}
\caption{\label{fig::triangular_lattice} A spanning cluster on the triangular lattice in a strip of vertical size $L_y=512$. The shortest path is in red and all the other sites belonging to the spanning cluster are in blue.}
\vspace*{-2mm}
\end{figure}

Let us consider critical site percolation on the triangular lattice, in a two-dimensional strip geometry of width $L_x$ and height $L_y$ ($L_y>L_x$), in units of lattice sites, see Fig. \ref{fig::triangular_lattice}. Each site is occupied with probability $p=p_c$. See Methods for details on the algorithm used to generate the curves. The largest cluster spans the lattice with non-zero probability, and the average shortest path length $\langle l \rangle$, defined as the number of sites in the path, scales as $\langle l \rangle \sim L_y^{d_\text{min}}$, where $d_\text{min}$ is the shortest path fractal dimension and its best estimation is $d_\text{min}=1.13077(2)$ \cite{Zhou12,Schrenk13}. There have been several attempts to compute exactly this fractal dimension \cite{Havlin84,Larsson87,Herrmann88,Tzschichholz89,Grassberger92b,Porto98,Deng10}. Most tentatives were based on scaling relations, conformal invariance, and Coulomb gas theory. But the existing conjectures have all been ruled out by precise numerical calculations. 
For example, Ziff computed the critical exponent $g_1$ of the scaling function of the pair-connectiveness function in percolation using conformal invariance arguments \cite{Ziff99b}. $g_1$ has been conjectured to be related to the fractal dimension of the shortest path \cite{Porto98}. In turn Deng \textit{et al.} conjectured a relation between $d_\text{min}$ and the Coulomb gas coupling for the random-cluster model \cite{Deng10}. Both conjectures were discarded by the latest numerical estimates of $d_\text{min}$ \cite{Grassberger99,Zhou12}. Thence, as recognized by Schramm in his list of open problems, a solid theory for the shortest path is still considered one of the major unresolved questions in percolation \cite{Schramm06}.

Impressive progress has recently been made in the field of critical lattice models using the Schramm-Loewner Evolution theory (SLE). In SLE, random critical curves are parametrized by a single parameter $\kappa$, related to the diffusivity of Brownian motion. Let us consider the case of a non self-touching curve, like the shortest path, defined in the upper half plane $\mathbb{H}$, that starts at the origin and grows towards infinity. Under a proper choice of parameters, it is possible to define a unique conformal map $g_t$ from $\mathbb{H} \setminus \gamma[0,t]$, i.e. the upper half-plane minus the curve $\gamma[0,t]$, onto $\mathbb{H}$ such that there exists a continuous real function $\xi_t$, and $g_t$ satisfies the stochastic Loewner differential equation,
\begin{equation}
\label{eq::chordalLoewnerEq}
\frac{\partial g_t(z)}{\partial t}=\frac{2}{g_t(z)-\xi_t},
\end{equation}
with $g_0(z)=z$. The function $\xi_t$ is called driving function. For details about the conformal map $g_t$ see Supplementary Information. We define \textit{chordal} $SLE_\kappa$ as the random collection of conformal maps in the upper-half plane that satisfy the Loewner equation with a driving function $\xi_t=\sqrt{\kappa}B_t$, where $B_t$ is a one-dimensional Brownian motion.

With the value of $\kappa$, one can obtain exactly several probability distributions for the curve, allowing to compute, for example, crossing probabilities and critical exponents \cite{Smirnov01,Lawler01,Smirnov01b}. SLE has been shown to describe many conformally invariant scaling limits of interfaces of two-dimensional critical models. In particular, $\mbox{SLE}_{6}$ has first been conjectured \cite{Schramm00} and later proved on the triangular lattice \cite{Smirnov01} to describe the hull in critical percolation \cite{Camia06}. SLE has been successfully used to compute rigorously other critical exponents of percolation-related objects \cite{Smirnov01b,Lawler01c} as, for example, the order parameter exponent $\beta$, the correlation length exponent $\nu$, and the susceptibility exponent $\gamma$ \cite{Smirnov01b}. More recently, the probability distributions of the hulls of the Ising model \cite{Coniglio89,Smirnov06,Smirnov10,Smirnov12} and of the Loop Erased Random Walks \cite{Majumdar92,Schramm00,Lawler04} were computed exactly. Therefore, it is legitimate to ask if the SLE techniques can help solving the long standing problem of the fractal dimension of the shortest path. 

Also, a possible description of the physical process through SLE gives interesting insights in new ways of generating the shortest path curves. Once $\mbox{SLE}_\kappa$ is established, the value of $\kappa$ suffices to generate, from only a Brownian motion, curves having the same statistical properties as the shortest path \cite{Kennedy09,Gherardi10,Miller12_preprint}. This can be very useful in the case of problems involving optimization processes like the shortest path, watersheds \cite{Daryaei12}, or spin glass problems \cite{Stevenson11,Bernard07,Amoruso06}, as traditional algorithms imply the exploration of large areas.

In this article, we will show that the numerical results are consistent with SLE predictions with $\kappa=1.04 \pm 0.02$. $SLE_\kappa$ curves have a fractal dimension $d_f$ related to $\kappa$ by $d_f=\min \left(2,1+\frac{\kappa}{8} \right)$ \cite{Beffara08}. From the estimate of the fractal dimension of the shortest path, one deduces the value of the diffusion coefficient $\kappa$ corresponding to an SLE curve of same fractal dimension; $\kappa_\text{fract}=1.0462 \pm 0.0002$. In what follows, we compute three different estimates of $\kappa$ using different analyses and compare them to $\kappa_\text{fract}$. In particular we consider the variance of the winding angle \cite{Duplantier88,Schramm00,Wieland03}, the left passage probability \cite{Schramm01}, and the statistics of the driving function \cite{Bernard06,Bernard07}. All estimates are in agreement with the one predicted from the fractal dimension, and therefore constitute a strong numerical evidence for the possibility of an SLE description of the shortest path.

\begin{figure}[t]
\vspace*{-5mm}
\begin{center}
\includegraphics[width=\columnwidth]{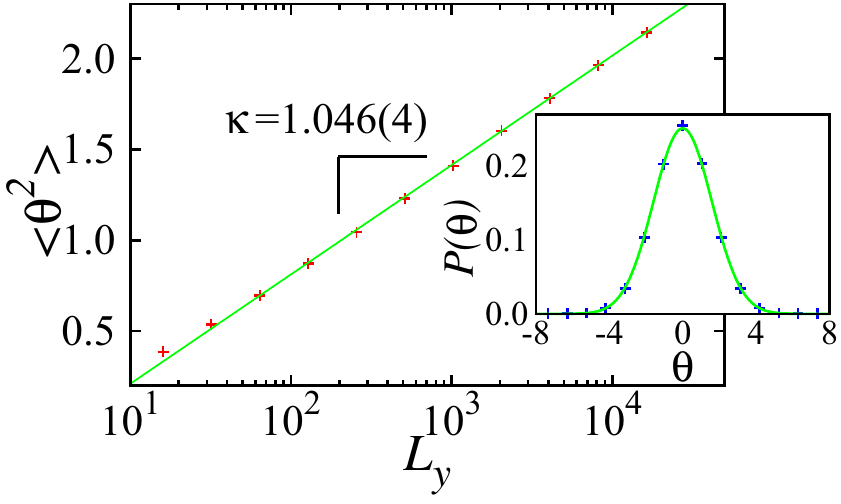}
\end{center}
\vspace*{-5mm}
\caption{\label{fig::windingAngle} Variance of the winding angle against the lattice size $L_y$. The analysis has been done for $L_y$ ranging from $16$ to $16384$. The statistics are computed over $10^4$ samples. The error bars are smaller than the symbol size. By fitting the results with Eq.~(\ref{eq::winding_angle}), one gets $\kappa_\text{winding}=1.046 \pm 0.004$. In the inset, the probability distribution of the winding angle along the curve is compared to the predicted Gaussian distribution, drawn in green, of variance $\frac{\kappa}{4}\ln(L_y)$ with $\kappa=1.046$ and $L_y=16384$.}
\vspace*{-2mm}
\end{figure}

\begin{figure*}[t]
\vspace*{-5mm}
\begin{center}
\includegraphics[width=0.6\paperwidth]{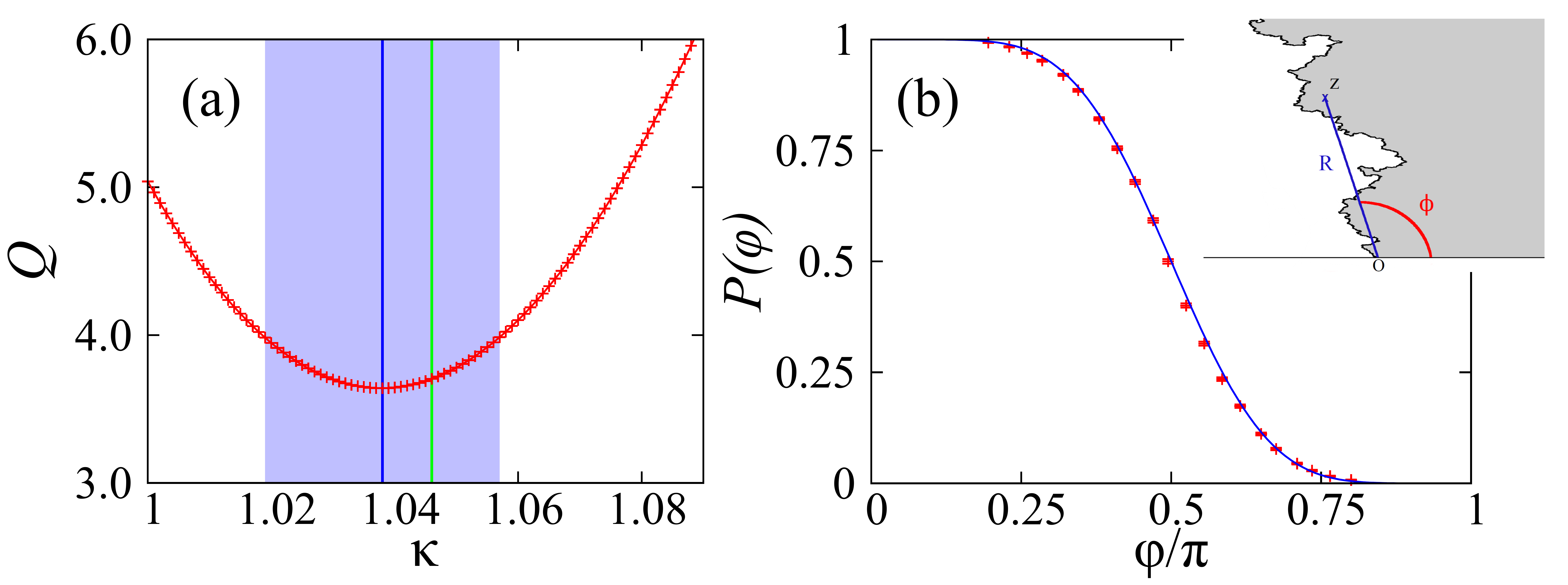}
\end{center}
\vspace*{-5mm}
\caption{\label{fig::lpp} Left passage probability test. (a) Weighted mean square deviation $Q(\kappa)$ as a function of $\kappa$, for $L_y=16384$. The vertical blue line corresponds the minimum of $Q(\kappa)$, and the green vertical line is a guide to the eye at $\kappa=\kappa_\text{fract}$. The minimum of the mean square deviation is at $\kappa_{LPP}=1.038\pm 0.019$. The light blue area corresponds to the error bar on the value of $\kappa_{LPP}$. We define the error bar $\Delta Q$ for the minimum of $Q(\kappa)$ using the fourth moment of the binomial distribution. The error $\Delta \kappa$ is defined such that $Q(\kappa \pm \Delta \kappa)-\Delta Q= Q(\kappa)+\Delta Q$. We considered 400 points, regularly spaced in $[-0.1L_x,0.1 L_x]\times [0.15 L_y, 0.35 L_y]$ which are then mapped through the inverse Schwarz-Christoffel mapping into $\mathbb{H}$ \cite{Driscoll02}. (b) Computed left passage probability as a function of $\phi/\pi$ for $R \in [0.70,0.75]$ and $\kappa=1.038$. The blue line is a guide to the eye of Schramm's formula (\ref{eq::lpp_shortest_path}) for $\kappa=1.038$.}
\end{figure*}

\section*{Results}

\textit{Winding angle.} The first result related to SLE deals with the winding angle. For each shortest path curve we have a discrete set of points $z_i$, called edges, on the lattice. The winding angle $\theta_i$ at each point $z_i$ can be computed iteratively as $\theta_{i+1}=\theta_i+\alpha_i$, where $\alpha_i$ is the turning angle between the two consecutive points $z_i$ and $z_{i+1}$. Duplantier and Saleur computed the probability distribution of the winding angle for random curves using conformal invariance and Coulomb gas techniques \cite{Duplantier88}. According to their result \cite{Schramm00}, for $SLE_{\kappa}$, the winding angle along all the edges of the curve exhibits a Gaussian distribution of variance
\begin{equation}
\label{eq::winding_angle}
\langle \theta^2 \rangle -\langle \theta \rangle ^2= b + \frac{\kappa}{4}\ln(L_y),
\end{equation}
where $b$ is a constant and $L_y$ is the vertical lattice size \cite{Wieland03}. Therefore, $\kappa/4$ corresponds to the slope of $\langle \theta^2 \rangle$ against $\ln(L_y)$. Figure \ref{fig::windingAngle} shows the results for the winding angle of the shortest path. The distribution is a Gaussian with a variance consistent with Eq.~(\ref{eq::winding_angle}). The estimate $\kappa_\text{winding}=1.046 \pm 0.004$ that we get from fitting the data with Eq.~(\ref{eq::winding_angle}) is in agreement with the value deduced from the fractal dimension.

\textit{Left passage probability.} In the following, we work with chordal SLE. Therefore, one has to conformally map the original curves into the upper half plane. This is done using an inverse Schwarz-Christoffel transformation (see Supplementary Information).

The shortest path splits the domain into two parts: the left and the right parts of the curve. The curve is said to pass at the left of a given point if this point belongs to the right side of the curve, see Fig.~\ref{fig::lpp}. For chordal $SLE_{\kappa}$ curves, Schramm has computed the probability of a curve to go to the left of a given point $z=R e^{i\phi}$, where $R$ and $\phi$ are the polar coordinates of $z$ \cite{Schramm01}. For a chordal $SLE_{\kappa}$ curve in $\mathbb{H}$, the probability $P_{\kappa}(\phi)$ that it passes to the left of $Re^{i\phi}$ depends only on $\phi$ and is given by Schramm's formula,
\begin{equation}
\label{eq::lpp_shortest_path}
P_{\kappa}(\phi)=\frac{1}{2}+\frac{\Gamma\left(4/\kappa\right)}{\sqrt{\pi}\Gamma\left(\frac{8-\kappa}{2\kappa}\right)}\cot(\phi)_2F_1\left(\frac{1}{2},\frac{4}{\kappa}, \frac{3}{2}, -\cot(\phi)^2\right),
\end{equation}
where $\Gamma$ is the Gamma function and $_2F_1$ is the Gauss hypergeometric function. We define a set of sample points $S$ in $\mathbb{H}$ for which we numerically compute the probability $P(z)$ that the curve passes to the left of these points. To estimate $\kappa$, we minimize the weighted mean square deviation $Q(\kappa)$ defined as,
\begin{equation}
\label{eq::lpp_quality_fit}
Q\left(\kappa\right)=\frac{1}{|S|}\sum_{z\in S} \frac{\left[P(z)-P_{\kappa}(\phi(z))\right]^2}{\Delta P(z)^2}, 
\end{equation}
where $|S|$ is the cardinality of the set $S$, and $\Delta P(z)^2$ is defined as $\Delta P(z)^2=\frac{P(z)(1-P(z))}{N_s-1}$, where $N_s$ is the number of samples \cite{Norrenbrock13}.

For a lattice size of $L_y=16384$, the minimum of the mean square deviation is observed for $\kappa_\text{LPP}=1.04 \pm 0.02$ as shown in Fig.~\ref{fig::lpp}. This value is in agreement with the estimate of $\kappa$ obtained from the fractal dimension and the winding angle.

\textit{Direct SLE.} The winding angle and left passage analyses are indirect measurements of $\kappa$. Therefore we also test the properties of the driving function directly in order to see if it corresponds to a Brownian motion with the expected value of $\kappa$.

\begin{figure}[t]
\vspace*{-5mm}
\begin{center}
\includegraphics[width=\columnwidth]{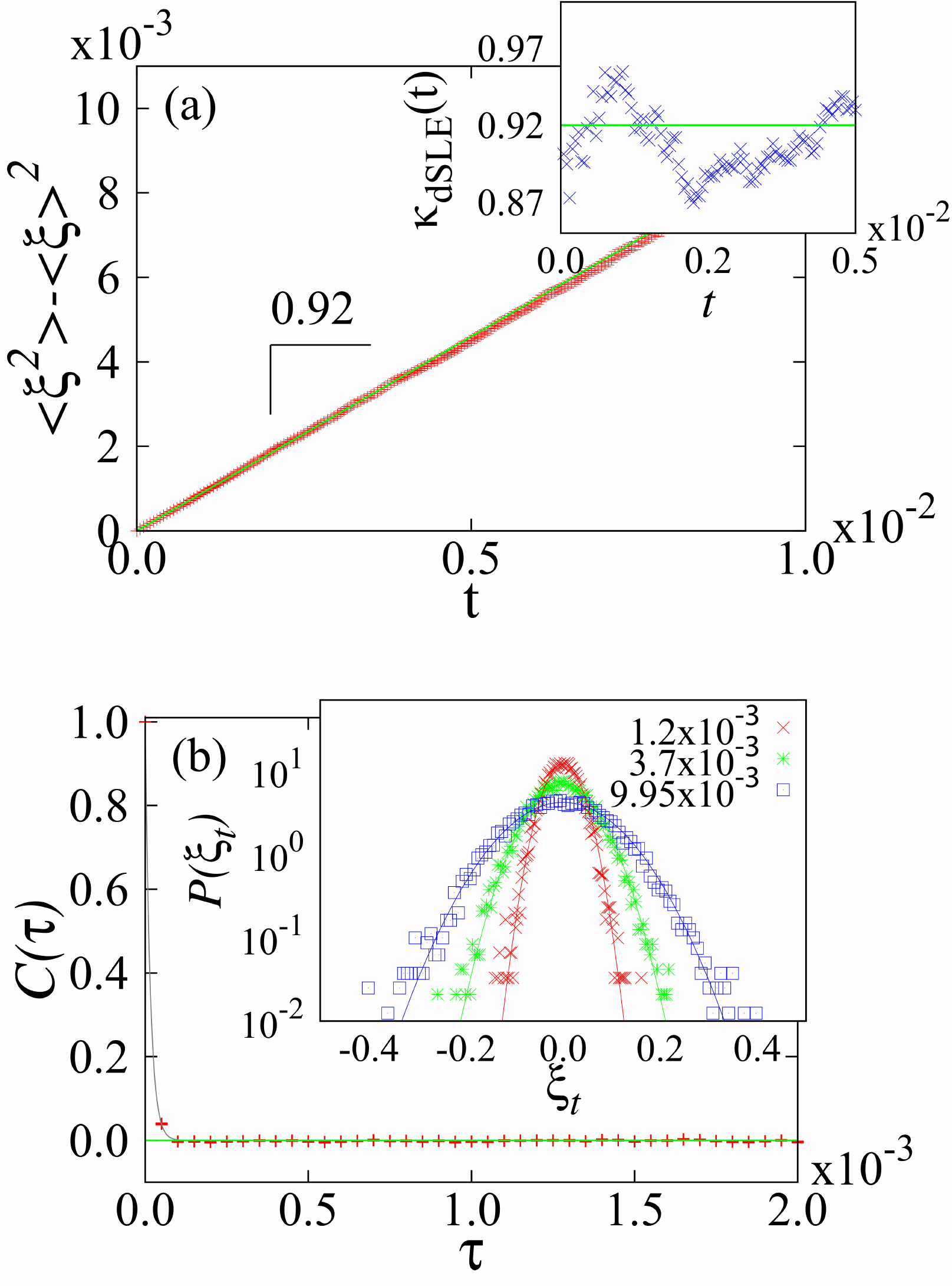}
\end{center}
\vspace*{-5mm}
\caption{\label{fig::directSLE} Driving function computed using the slit map algorithm. (a) Mean square deviation of the driving function $\langle \xi_t^2 \rangle$ as a function of the Loewner time $t$. The diffusion coefficient $\kappa$ is given by the slope of the curve. In the inset we see the local slope $\kappa_\text{dSLE}(t)$. The thick green line is a guide to the eye corresponding to $\kappa_{\text{dSLE}}=0.92$. (b) Plot of the correlation $C(t,\tau)$ given by Eq.~(\ref{eq::autocorrelation}), and averaged over 50 time steps. The averaged value is denoted $C(\tau)$. In the inset are shown the probability distributions of the driving function for three different Loewner times $t_1=1.2\times 10^{-3}$, $t_2=3.7\times 10^{-3}$ and $t_3=9.95\times 10^{-3}$. The solid lines are guides to the eye of the form $P(\xi_t)=\frac{1}{\sqrt{2 \pi \kappa t_i}}\exp\left(-\frac{\xi_t^2}{2 \kappa t_i}\right)$, for $i=1,2,3$.}
\vspace*{-5mm}
\end{figure}

As for the left passage probability, we consider the chordal curves in the upper half plane, starting at the origin and growing towards infinity. We want to compute the driving function $\xi_t$ underlying the process. For that, we numerically solve Eq.~(\ref{eq::chordalLoewnerEq}) by considering the driving function to be constant within a small time interval $\delta t$, thus one obtains the slit map equation \cite{Cardy05,Kennedy09},
\begin{equation}
\label{eq::chordalSlitMap}
g_t(z)=\xi_t + \sqrt{\left(z-\xi_t\right)^2 + 4\delta t}.
\end{equation}
We start with $\xi_t=0$ at $t=0$ and the initial points of the curve $\{z_0^0=0, z_1^0=z_1, \ldots, z_N^0=z_N \}$, and map recursively all the points $\{z_{i}^{i-1}, \ldots , z_N^{i-1}\}$, $i>0$, of the curve to the points $\{z_{i+1}^{i}=g_{t_i}(z_{i+1}^{i-1}), \ldots, z_N^{i}=g_{t_i}(z_N^{i-1})\}$ through the map $g_{t_i}$, sending $z_{i}^{i-1}$ to the real axis by setting $\xi_{t_i}=\mbox{Re} \{ z_i^{i-1} \}$ and $\delta t_i=t_i-t_{i-1}=\left(\mbox{Im} \{z_i^{i-1}\} \right)^2/4$ in Eq. (\ref{eq::chordalSlitMap}). $\mbox{Re}\{\}$ and $\mbox{Im}\{\}$ are respectively the real and imaginary parts. In the case of $SLE_{\kappa}$ the extracted driving function gives a Brownian motion of variance $\kappa$. The direct SLE test consists in verifying that the driving function is a Brownian motion and compute its variance $\langle \xi_t^2\rangle - \langle \xi_t \rangle ^2$ to obtain the value of $\kappa$. The variance should behave as $\langle \xi_t^2 \rangle- \langle \xi_t \rangle ^2=\kappa t$.

We extract the driving function $\xi_t$ of the shortest path curves using the slit map, Eq.~(\ref{eq::chordalSlitMap}). Figure \ref{fig::directSLE} shows the variance of the driving function as a function of the Loewner time $t$. We observe a linear scaling of the variance with $t$. The local slope $\kappa_{\text{dSLE}}(t)$ is shown in the inset of Fig.~\ref{fig::directSLE}a. In Fig.~\ref{fig::directSLE}b, we plot the mean correlation function $C(\tau)=\langle C(t,\tau) \rangle _t$ of the increments $\delta \xi_t$ of the driving function, where the correlation function is defined as,
\begin{equation}
\label{eq::autocorrelation}
C(t, \tau)=\frac{\langle \delta \xi_{t+\tau} \delta \xi_t \rangle - \langle \delta \xi_{t+\tau}\rangle \langle \delta \xi_t \rangle}{\sqrt{\left(\langle \delta \xi_{t+\tau}^2\rangle-\langle \delta \xi_{t+\tau} \rangle ^2\right)\left(\langle \delta \xi_t^2\rangle-\langle \delta \xi_t \rangle ^2\right)}}.
\end{equation} 
One sees that the correlation function vanishes after a few time steps. The initial decay is due to the finite lattice spacing, which introduces short range correlations. But in the continuum limit, the process is Markovian, with a correlation function dropping immediately to zero. In the inset of Fig. \ref{fig::directSLE}b, we show the probability distribution of the increments for different $t$. This distribution is well fitted by a Gaussian, in agreement with the hypothesis of a Brownian driving function. From this result and the estimates of the diffusion coefficient computed for several lattice sizes, we obtain $\kappa=0.9 \pm 0.2$.

We note that the numerical results obtained with the direct SLE method are less precise than with the other analyses and, therefore, characterized by larger error bars, as is well known in the literature \cite{Bernard06,Bernard07,Bernard07b,Bogomolny07,Daryaei12,Najafi12}. The result we have obtained for $\kappa$ is in agreement with the ones obtained with the fractal dimension, winding angle, and left-passage probability.

We also extracted the driving function of the curves in dipolar space, i.e. defining the curves as starting from the origin and growing in the strip (see Supplementary Information). We also obtained a value of $\kappa$ consistent with the fractal dimension.

\section*{Discussion}

All tests are consistent with SLE predictions. The numerical results obtained with the winding angle, left-passage, and direct SLE analyses are in agreement with the latest value of the fractal dimension. Being SLE implies that the shortest path fulfills two properties: conformal invariance and domain Markov property (DMP). Thus, the agreement with SLE predictions lends strong arguments in favor of conformal invariance and DMP of the shortest path.

The DMP is related to the evolution of the curve in the domain of definition. Let us consider the shortest path $\gamma$ defined in a domain $\mathbb{D}$, starting in $a$ and ending in $b$. We take a point $c$ on the shortest path different from $a$ and $b$. Then if the DMP holds, one would have that
\begin{equation}
\label{eq::Domain_Markov_property}
\mathbb{P}_{\mathbb{D}}\left(\gamma[a,b] | \gamma[a,c]\right)=\mathbb{P}_{\mathbb{D}\setminus \gamma[a,c]}\left(\gamma[c,b]\right),
\end{equation}
where $\gamma[c,b]$ is the shortest path starting in $c$ and ending in $b$ in the domain $\mathbb{D}$ except the curve $\gamma[a,c]$, denoted as $\mathbb{D} \setminus \gamma[a,c]$, and $\mathbb{P}_\mathbb{D}$ and $\mathbb{P}_{\mathbb{D} \setminus \gamma[a,c]}$ are the probabilities in the domains $\mathbb{D}$ and $\mathbb{D} \setminus \gamma[a,c]$ respectively. One can classify the models as the ones for which DMP holds already on the lattice, and the ones for which it holds only in the scaling limit. Many classical models, like the percolation hulls, the LERW, or the Ising model \cite{Bauer06} for example, belong to the first case. But some two-dimensional spin glass models with quenched disorder \cite{Bernard07,Stevenson11} are believed to only fulfill DMP in the scaling limit. Our numerical results suggest that, for the shortest path, DMP holds at least in the scaling limit. Further studies should be done to test the validity of DMP on finite lattices.

The second result we can expect if SLE is established for the shortest path is conformal invariance. Conformal invariance, being a powerful tool to compute critical exponents, is of interest for the study of the shortest path. Conformal invariance, associated to Coulomb gas theory for example, could be useful to develop a field theoretical approach of the shortest path. There is no proof of conformal invariance of the shortest path, but our numerical results give strong support to this hypothesis. For example, the expression of the winding angle is based on conformal invariance and agrees with the predictions based on the fractal dimension. Also the left passage probabilities and the direct SLE measurements have been performed on curves conformally mapped to the upper half plane and gave consistent results. In addition, we obtained the same estimate of $\kappa$ by extracting the driving function in chordal and dipolar space. However, even if the scaling limit would not be conformally invariant, our results suggest that one could still apply SLE techniques to the study of this problem, as some SLE techniques have also been used to study off-critical and especially non conformal problems \cite{Bauer08, Nolin09, Bauer09, Makarov09, Garban13}.

Analyzing the shortest path in terms of an SLE process would give a deeper understanding of probability distributions of the shortest path, allowing to compute more quantities, like for example the hitting probability distribution of the shortest path on the upper boundary segment \cite{Bauer05}.

\section*{Methods}
We generate random site percolation configurations on a rectangular lattice $L_x \times L_y$ with triangular mesh, where $L_x$ and $L_y$ are respectively the horizontal and vertical lattice sizes, in units of lattice sites. The sites of the lattice are occupied randomly with the critical probability $p_c=\frac{1}{2}$. If the configuration percolates, we obtain the spanning cluster and identify the shortest path between the top and bottom layers using a burning method \cite{Herrmann84,Stauffer94,Grassberger99b}. In short, we burn the spanning cluster from the bottom sites, indexing the sites by the first time they have been reached, and stop the burning when we reach for the first time the top line. We then start a second burning from the sites on the top line that have been reached by the first burning, burning only sites with lower index. With this procedure, we identify all shortest paths from the bottom line to the top one. We randomly choose with uniform probability one of these paths. The results presented in the paper are for $L_y$ ranging from $16$ to $16384$ and an aspect ratio of $L_x/L_y=1/2$. We generated 10000 samples and discarded the paths touching the vertical borders.

\begin{acknowledgments}
\textbf{Acknowledgments.} The authors would like to thank W. Werner and E. Daryaei for helpful discussions. We acknowledge financial support from the ETH Risk Center, the
Brazilian institute INCT-SC, and grant number FP7-319968 of the
European Research Council.
\end{acknowledgments}

\appendix

\section*{Appendix}

\textit{From dipolar to chordal curves.} The curves we generate with the algorithm described above are defined in a stripe starting at the bottom boundary and ending at the upper one. However, we use results that are valid for \textit{chordal} curves,  like the left-passage probability formula computed by Schramm \cite{Schramm01}. \textit{Chordal} curves are defined in the upper half plane, starting at the origin and growing towards infinity. Therefore we have to map conformally the original curves into the upper half plane by an inverse Schwarz-Christoffel transformation \cite{Driscoll02}. Our curves are generated with ``free'' boundary conditions, i.e. without any constraints on the boundaries, such that the shortest path has no fixed starting and ending points. We relocate the curves in order for them to start at the origin; the curves are now defined in the rectangle $[-L_x,L_x]\times [0,L_y]$ in lattice units. We then use an inverse Schwarz-Christoffel transformation that maps the rectangle $[-L_x,L_x]\times [0,2L_y]$ into the upper half plane with the point $(0,2L_y)$ being mapped to infinity.

\textit{Loewner's equation in chordal space.} Let us consider the case of a simple, i.e. non self-touching, \textit{chordal} curve $\gamma(t)$ defined in the upper half plane $\mathbb{H}$. From the Riemann mapping theorem, there exist conformal maps $g_t$ from the upper half plane minus the curve $\gamma[0,t]$, denoted as $\mathbb{H}\setminus \gamma[0,t]$, onto $\mathbb{H}$ such that $g_t(\infty)=\infty$. If $g_t$ is such a map, then all the conformal maps from $\mathbb{H}\setminus \gamma[0,t]$ onto $\mathbb{H}$ such that $g_t(\infty)=\infty$ are of the form $\alpha g_t+ \beta$, with $\alpha >0$ and $\beta \in \mathbb{R}$. In order to fix uniquely the map, one has to choose the dilatation and translation factors $\alpha$ and $\beta$. This is done by the following ``hydrodynamical'' normalization: the map is chosen such that $\lim_{z\rightarrow \infty} g_t(z)-z=0$.

We parametrize the curve such that ${\lim_{z \rightarrow \infty}z(g_t(z)-z)=2t}$. Then, there exists a continuous real function $\xi_t$ such that $g_t$ satisfies the stochastic Loewner differential equation,
\begin{equation}
\label{eq::chordalLoewnerEq}
\frac{\partial g_t(z)}{\partial t}=\frac{2}{g_t(z)-\xi_t} \mbox{, and } g_0(z)=z.
\end{equation}
The function $\xi_t$ is called driving function. It corresponds to the tip of the curve mapped by $g_t$ to the real axis $\xi_t=g_t(\gamma(t))$ and $\xi_0=0$. We define \textit{chordal} SLE as the random collection of conformal maps in the upper-half plane that satisfy Loewner's equation with a driving function $\xi_t=\sqrt{\kappa}B_t$, where $B_t$ is a one-dimensional Brownian motion starting at the origin.

\textit{Direct SLE in dipolar space.} 
\begin{figure}[t]
\vspace*{0mm}
\begin{center}
\includegraphics[width=0.7\columnwidth]{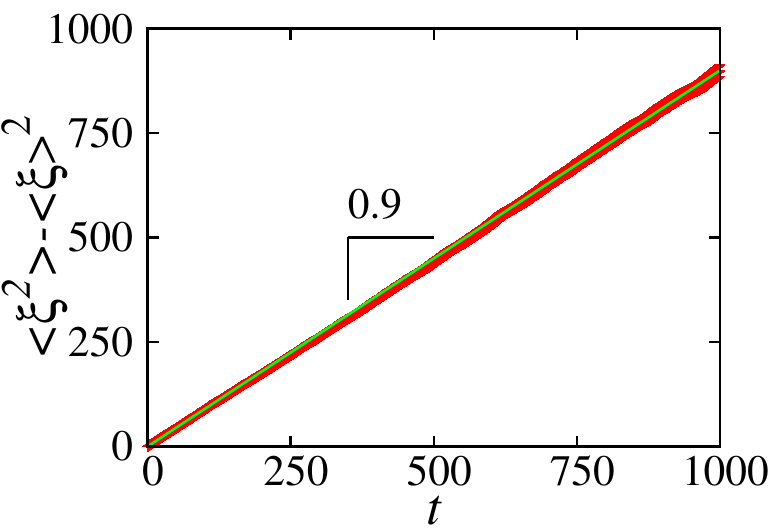}
\end{center}
\vspace*{-5mm}
\caption{\label{fig::directSLE_dipolar} Driving function computed using the dipolar slit map given by Eq.~(\ref{eq::dipolar_slit_map}), for $L_y=16384$. The mean square deviation of the driving function $\langle \xi_t^2 \rangle-\langle \xi_t \rangle^2$ is plotted as a function of the Loewner time $t$. The diffusion coefficient $\kappa$ is given by the slope of the curve.}
\end{figure}
Let us consider the case of a \textit{dipolar} curve growing in the strip $\mathbb{S}$ of height $L_y$, that starts at the origin and stops the first time it hits the upper boundary. We again study the case of simple curves. By the Riemann mapping theorem, there exist conformal maps from the strip $\mathbb{S}$ minus the curve $\gamma[0,t]$ into the strip $\mathbb{S}$ such that $g_t(\infty)=\infty$ and $g_t(-\infty)=-\infty$. The map $g_t$ is then defined up to a translation by a real constant. It is made unique by choosing the normalization $\lim_{z \rightarrow \infty} g_t(z)+g_t(-z)=0$. One parametrizes the curves such that $\lim_{z \rightarrow \infty} g_t(z)-z=t$, where $t$ is called the Loewner time. \textit{Dipolar} SLE is defined as the collection of conformal maps $g_t$ satisfying the following stochastic differential equation
\begin{equation}
\label{eq::dipolarLoewnerEq}
\frac{\partial g_t(z)}{\partial t}=\frac{\pi/L_y}{\tanh\left(\pi\left(g_t(z)-\xi_t\right)/2L_y\right)} \mbox{, and } g_0(z)=z,
\end{equation}
where $\xi_t=\sqrt{\kappa}B_t$ and $B_t$ is a one dimensional Brownian motion starting at the origin \cite{Bauer05,Bauer09}. 

Using the theory of dipolar SLE, one can develop a numerical method to compute the driving function of dipolar curves, as has been done in the main text for chordal curves. Therefore one has to solve Eq.~(\ref{eq::dipolarLoewnerEq}). Consider a dipolar curve defined by the initial set of points $\{z_0^{0},...,z_N^{0}\}$. One maps recursively the sequence of points $\{z_i^{i-1},...,z_N^{i-1}\}$ of the mapped curve to the shortened sequence $\{z_{i+1}^{i},...,z_N^{i}\}$ by the conformal map
\begin{equation}
\label{eq::dipolar_slit_map}
g_{t_i}(z)=\xi_{t_i}+2 \frac{L_y}{\pi} \cosh^{-1}\left(\frac{\cosh\left(\pi(z-\xi_{t_i})/2L_y\right)}{\cos(\Delta_i)}\right),
\end{equation}
where ${\xi_{t_i}=\mbox{Re}\{z_i^{i-1}\}}$ and ${\delta_{t_i}=t_i-t_{i-1}}=-2(L_y/\pi)^2\ln\left(\cos(\Delta_i)\right)$, with $\Delta_i=\pi\mbox{Im}\{z_i^{i-1}\}/2L_y$ and $\mbox{Re}\{\}$ and $\mbox{Im}\{\}$ being respectively the real and imaginary parts \cite{Bernard07}. If the curve follows $\mbox{SLE}_{\kappa}$ statistics, then the driving function is a one dimensional Brownian motion of variance $\langle \xi_t^2 \rangle-\langle \xi_t \rangle^2=\kappa t$. In Fig. (\ref{fig::directSLE_dipolar}) we show the mean variance, average over different curves, of the driving function of dipolar shortest path curves as a function of the Loewner time. One obtains a value of $\kappa$ in agreement with the value obtained in the chordal case $\kappa=0.9 \pm 0.2$. 


\end{document}